\documentclass{IEEEtran4PSCC}
\ifCLASSINFOpdf
   \usepackage[pdftex]{graphicx}
\else
   \usepackage[dvips]{graphicx}
\fi
\usepackage[cmex10]{amsmath}
\usepackage{pdfpages}
\usepackage{graphicx}
\usepackage{epsfig}
\usepackage{siunitx}
\usepackage{rotating}
\usepackage{tikz}
\usepackage[oldvoltagedirection]{circuitikz}
\usepackage{wrapfig}

\usepackage{algorithm}
\usepackage{algpseudocode}
\usepackage{amsmath}
\usepackage{amsmath,amsfonts,amssymb}
\usepackage{graphicx}
\usepackage[font=footnotesize]{caption}
\usepackage{subcaption}
\usepackage{float}

\usepackage{nomencl}
\usepackage{multicol}
\usepackage{etoolbox}

\usepackage{booktabs}
\usepackage{svg}

\makeatletter
\let\old@ps@headings\ps@headings
\let\old@ps@IEEEtitlepagestyle\ps@IEEEtitlepagestyle
\def\psccfooter#1{%
    \def\ps@headings{%
        \old@ps@headings%
        \def\@oddfoot{\strut\hfill#1\hfill\strut}%
        \def\@evenfoot{\strut\hfill#1\hfill\strut}%
    }%
    \def\ps@IEEEtitlepagestyle{%
        \old@ps@IEEEtitlepagestyle%
        \def\@oddfoot{\strut\hfill#1\hfill\strut}%
        \def\@evenfoot{\strut\hfill#1\hfill\strut}%
    }%
    \ps@headings%
}
\makeatother

\psccfooter{%
        \parbox{\textwidth}{\hrulefill \\ \small{24th Power Systems Computation Conference} \hfill \begin{minipage}{0.2\textwidth}\centering \vspace*{4pt} \includegraphics[scale=0.06]{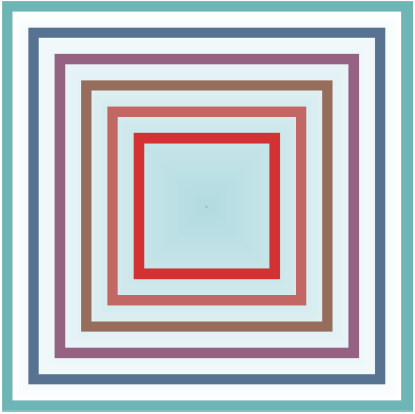}\\\small{PSCC 2026} \end{minipage} \hfill \text{ \ \ \ }\text{ \ \ }\small{  Limassol, Cyprus --- June 8-12, 2026}}%
}

\begin{document}
%
\title{\vspace{7mm}Learning a Non-linear Surrogate Model for Multistage Stochastic Transmission Planning}



\author{\IEEEauthorblockN{Victor Schmitt\IEEEauthorrefmark{1},
Farzaneh Pourahmadi\IEEEauthorrefmark{1},
Angela Flores-Quiroz\IEEEauthorrefmark{2}, 
Pablo Apablaza\IEEEauthorrefmark{3},
Pierluigi Mancarella\IEEEauthorrefmark{3}\IEEEauthorrefmark{4}}
\IEEEauthorblockA{\IEEEauthorrefmark{1} Department of Wind and Energy Systems,
Technical University of Denmark, Kgs. Lyngby, Denmark}
\IEEEauthorblockA{\IEEEauthorrefmark{2} Department of Electrical Engineering, University of Chile, Chile}
\IEEEauthorblockA{\IEEEauthorrefmark{3} Department of Electrical and Electronic Engineering,
The University of Melbourne, Melbourne, Australia}
\IEEEauthorblockA{\IEEEauthorrefmark{4}School of Electrical and Electronic Engineering, The University of Manchester, Sackville Street, Manchester M13 9PL, UK}
}

\maketitle

\begin{abstract}
Transmission expansion planning (TEP) plays a critical role in ensuring power system reliability and facilitating the integration of renewable energy resources. However, this process requires planners to constantly deal with significant uncertainty. While multistage stochastic TEP models provide a robust framework for identifying investment plans under uncertainty, the rapid growth in problem size hinders their computational tractability. To address this challenge, this paper develops a hybrid machine learning-optimisation framework for stochastic TEP. The proposed approach uses investment decisions and uncertainty scenarios as input features to train surrogate neural networks, which are then reformulated as mixed-integer linear constraints and embedded within an optimisation model. The surrogate model approximates expected operational costs to inform TEP decisions, reducing the burden arising from large operational problems. Case study applications on IEEE test systems demonstrate that, after training, the proposed approach achieves near-optimal investment costs while reducing total computational time by up to a factor of around 13 compared to a single full-optimisation stochastic formulation. This enables performing extensive multi-scenario analysis and stress testing that would otherwise be computationally prohibitive at scale.
\end{abstract}

\begin{IEEEkeywords}
Machine Learning, Neural Networks, Non-linear Surrogate Model, Power System Planning, Multistage Stochastic Transmission Expansion Planning
\end{IEEEkeywords}


\section*{Nomenclature}
\begin{IEEEdescription}[\IEEEusemathlabelsep\IEEEsetlabelwidth{$\text{net..}$}]
\item[\textbf{Parameters}]
\item[$b_{m-1}$] Vector of biases, layer $m$
\item[$c_g$]{Generation cost, generator $g$ [\$/MW]}
\item[$c_{l,s}^{\text{ann}}$]{Present value of annuitised investment costs, discounted over the relevant planning horizon starting from the year of investment, candidate line $l$ [\$]}
\item[$D_{b,t,s}$]{Load, bus $b$, time $t$, node $s$ [MW]}
\item[$\Bar{F}_l$]{Max. flow limit, line $l$ [MW]}
\item[$N$] Number of training runs
\item[$\Bar{P}_g$]{Max. capacity, generator $g$ [MW]}
\item[$r$]{Discount rate [-]}
\item[$\text{d}_{s}$] Demand time series, node $s$
\item[$\text{s}_{s}$] Solar generation time series, node $s$
\item[$\text{w}_{s}$] Wind generation time series, node $s$
\item[VoLL]{Value of Lost Load [\$/MWh]}
\item[$W_{m-1}$] Matrix of weights, layer $m$
\item[$X_{s}$] Matrix of input parameters, node $s$
\item[year$_s$]{Year, node $s$}

\item[$\alpha$] Initial learning rate
\item[$\delta$] Huber loss delta
\item[$\gamma$] Power system reliability standard
\item[$\kappa$] Scaling factor
\item[$\lambda$] Regularisation coefficient
\item[$\pi_{s}$] Probability of node $s$
\item[$\theta$] Neural network parameters
\item[$\upsilon$] Units per hidden layer
\item[$\boldsymbol{\tau}$] Transmission line parameters
\item[$\ell$] Number of hidden layers
\end{IEEEdescription}

\begin{IEEEdescription}[\IEEEusemathlabelsep\IEEEsetlabelwidth{$LS_{s}^{\text{tot}}$}]
\item[\textbf{Decision variables and functions}]
\item[$C_{s}^{\text{gen-tot}}$] Total generation costs, node $s$ [\$]
\item[$f_{\theta}$] Surrogate model function
\item[$\boldsymbol{f}_{l,t,s}$]{Power flow, line $l$, time $t$, node $s$ [MW]}
\item[$h_{m}$] Neural network output of layer $m$
\item[$LS_{s}^{\text{tot}}$] Load shedding, node $s$ [MWh]
\item[$\boldsymbol{LS}_{b,t,s}$]{Load shedding, bus $b$, time $t$, node $s$ [MW]}
\item[$\boldsymbol{p}_{g,t,s}$]{Power generation, generator $g$, time $t$, node $s$ [MW]}
\item[$\hat{\boldsymbol{C}}_{s}^{\text{gen-tot}}$] Predicted total generation costs, node $s$ [\$]
\item[$\hat{\boldsymbol{LS}}_{s}^{\text{tot}}$] Predicted load shedding, node $s$ [MWh]
\item[$\tilde{\boldsymbol{y}}_{l,s}$]{Stage-specific investment decision, candidate line $l$, node $s$}
\item[$\boldsymbol{y}_{l,s}^{\text{built}}$]{Operational status variable indicating whether candidate line $l$ is operational, node $s$}
\item[$\boldsymbol{y}_{l,s}^{\text{inv}}$]{Investment decision variable indicating whether candidate line $l$ is selected for investment, node $s$}

\item[$\sigma(\cdot)$] ReLU activation function
\end{IEEEdescription}

\begin{IEEEdescription}[\IEEEusemathlabelsep\IEEEsetlabelwidth{$e\in\mathcal{E}$}]
\item[\textbf{Indices and sets}]
\item[$\mathcal{D}_m$] Set of neural network nodes, layer $m$
\item[$g\in\mathcal{G}$]{Set of generators $g$}
\item[$l\in\mathcal{L}$]{Set of all transmission lines $l$}
\item[$l\in\mathcal{L}^{\text{C}}$] Set of candidate lines
\item[$l\in\mathcal{L}_{\overline{b}}$]{Lines leaving bus $b$}
\item[$l\in\mathcal{L}_{\underline{b}}$]{Lines entering bus $b$}
\item[$m\in\mathcal{M}$] Set of neural network layers
\item[$\mathcal{P}(s)$] Parent node of node $s$
\item[$s\in\mathcal{S}$] Nodes in the scenario tree
\item[$s\in\mathcal{S}_{T}$] Nodes in the scenario tree for a given stage $T$
\item[$t\in T$]{Set of time periods $t$}

\item[$\Gamma$] Set of possible network configurations
\end{IEEEdescription}

\section{Introduction}
Transmission infrastructure plays a critical role in modern electric power systems, particularly by supporting system reliability and cost efficiency through the integration and transport of large shares of renewable energy. However, significant uncertainties in system development and other factors (e.g., load growth, fuel prices, uptake of renewables) pose major challenges for decision-making in transmission planning.





Commonly employed deterministic methods for TEP are prone to producing suboptimal investment plans under uncertain future conditions \cite{moreno2017}. Stochastic transmission expansion planning (STEP) frameworks have gained prominence in the technical literature for explicitly accounting for uncertainty, thereby producing more robust investment strategies \cite{spyrou2024}. STEP frameworks aim to minimise the total expected cost of investment and operational decisions across a set of scenarios aggregated into a decision tree.  However, expanding scenario trees to include diverse system development pathways and the need for higher operational fidelity increases the computational burden, rendering STEP models intractable and therefore not widely applicable in realistic settings.




Supervised machine learning (ML) methods offer a promising alternative for addressing computational scalability challenges. While ML has been widely applied to scenario reduction \cite{fast_zhan_2017} or to accelerate the solution of the AC optimal power flow \cite{zamzam2020}, its potential for long-term planning problems remains largely unexplored. This paper investigates how a learning-based approach could alleviate the computational burden of TEP problems. We propose using a neural network-based surrogate model that approximates the operational cost component of a multistage stochastic expansion planning model. Leveraging the decision tree structure of STEP models, neural networks (NNs) with Rectified Linear Unit (ReLU) activation functions are reformulated as sets of linear constraints, enabling their direct embedding into an optimisation-based STEP model. This approach preserves the structure of the original optimisation framework while significantly improving computational tractability. 
\vspace{-1mm}
\subsection{Literature Review}
Planners typically consider a multi-year horizon, requiring the development of long-term planning strategies that determine what to invest in as well as where and when to procure. This multi-year temporal dimension significantly increases problem complexity by introducing numerous investment and operational variables and constraints, thereby demanding substantial computational resources, particularly for large-scale systems. 


To overcome the structural limitations of deterministic approaches in accounting for long-term uncertainty, recent work on multistage STEP models highlights their suitability in revealing the value of flexible planning strategies \cite{moreno2017}. For example, \cite{apablaza2024} developed a stochastic framework to assess the impact of distributed energy resources on expansion planning, showing that stochastic modelling could produce more robust and coherent investment strategies than deterministic methods.

However, incorporating detailed operational flexibility while considering the full planning horizon remains computationally demanding and often intractable in real-world applications. As problem size increases, memory requirements grow substantially, leading to slower and more expensive computations, which are impractical when planners must run multiple model instances. To mitigate this, \cite{flores2016} introduced a column generation (CG) approach for generation expansion planning, iteratively adding only the most promising variables to a restricted master problem and thereby reducing computational burden. Building on this, \cite{flores2021} proposed a multistage stochastic planning framework that jointly considers unit commitment, energy storage, reserve allocation, and transmission investments. By applying Dantzig–Wolfe (DW) decomposition and CG with column sharing across subproblems, their study reported up to an 86\% improvement in computational efficiency. Nonetheless, DW decomposition is in some cases susceptible to tailing-off effects and convergence issues as the size of the master problem increases, which could limit its applicability \cite{cg_book}.

More recently, researchers have turned to investigate the merits of ML techniques in alleviating the computational burden inherent in large-scale stochastic planning. For example, Benders' decomposition can be well-suited for two-stage STEP problems, as it separates investment and operational decisions. However, the master problem often grows rapidly when aggregating binary variables from subproblems, limiting tractability. To address this, \cite{jia2019} proposed a Support Vector Machine-based approach to identify representative optimality cuts, selectively adding only those that enhance convergence. This idea was further extended by the authors of \cite{borozan2024, Borozan_2025}, who employed Random Forests to predict effective versus redundant cuts, thereby improving both upper and lower bound estimates. Despite these advances, classical Benders decomposition may not be suitable for multistage settings when integer investment variables are considered, as the dual formulation is not well defined \cite{philpott2009dantzig}. Other studies have explored the potential of neural network and surrogate-based approaches, highlighting their ability to reduce computational cost without sacrificing accuracy. For instance, the authors in \cite{LEDEE2025125812} develop ML-based surrogate models for multi-energy system design, demonstrating that careful choices in model architecture, sampling, and scaling could improve predictive accuracy under limited data. Similarly, \cite{10267731} proposes a surrogate-based model for decoupling transmission and distribution planning, capturing distribution-level flexibility while achieving computational speed-ups of up to two orders of magnitude. Finally, \cite{KeimMachineLC} proposes a complementary approach that uses ML to accelerate the solution of subproblems within the CG method.

Beyond power systems research, the \textit{Neur2SP} framework \cite{dumouchelle2022neur2spneuraltwostagestochastic} employs NN surrogates for two-stage stochastic optimisation problems. Our work extends this approach to multistage STEP, introducing a scenario node-decomposed surrogate structure embedded within the optimisation problem that directly approximates expected operational costs and load shedding. This enables scalability to larger systems and deeper scenario trees without necessarily relying on standard decomposition techniques. 

While ML has been increasingly explored to support subtasks in STEP, such as scenario reduction or accelerating decomposition methods, its use as a standalone surrogate to directly assist in solving the full optimisation problem remains largely unexplored. This study contributes to this emerging area by proposing a systematic framework for generating training data and designing NN architectures tailored to TEP applications. It further develops a novel multistage STEP formulation in which operational costs are approximated with ReLU NNs reformulated as mixed-integer linear constraints to reduce computational complexity. 



The remainder of this paper is structured as follows. Section II presents the optimisation-based formulation for the multistage STEP model. Section III describes the methodology for data sampling and NN training, and embedding into an optimisation-based STEP framework. Section IV presents the results and discussion. Section V concludes and outlines directions for future research. 

\section{Optimisation model for the multistage stochastic transmission planning problem}
\label{sec:model_sto}

The optimisation model for solving the multistage STEP problem is compactly formulated via eqs. (\ref{eq:objective_sto_ms_disc_tot})-(\ref{eq:unique_inv_tot}). 

{\footnotesize
\begin{align}
&\min_{\substack{\tilde{\boldsymbol{y}}_{l,s}, \boldsymbol{y}_{l,s}^{\text{inv}}, \boldsymbol{y}_{l,s}^{\text{built}}, \\
    \boldsymbol{p}_{g,t,s}, \boldsymbol{f}_{l,t,s}, \boldsymbol{LS}_{b,t,s}}}  
\sum_{s\in\mathcal{S}}\sum_{l\in\mathcal{L}^C}
\pi_s c^{\text{ann}}_{l,s}\tilde{\boldsymbol{y}}_{l,s} + && \notag\\
&\sum_{s\in\mathcal{S}}\frac{\pi_s}{(1+r)^{\text{year}_s}}
\left(
\sum_{g\in\mathcal{G}}\sum_{t\in T} c_g \boldsymbol{p}_{g,t,s}
+
\sum_{b\in\mathcal{B}}\sum_{t\in T} \text{VoLL}\,\boldsymbol{LS}_{b,t,s}
\right)
\label{eq:objective_sto_ms_disc_tot}
\end{align}
}
\vspace{-4mm}
\begin{align}
\text{s.t.}\quad
& \sum_{g\in\mathcal{G}_{(b)}} \boldsymbol{p}_{g,t,s} + \boldsymbol{LS}_{b,t,s} - &&
\notag \\
& \sum_{l\in\mathcal{L}_{\overline b}} \boldsymbol{f}_{l,t,s}
+ \sum_{l\in\mathcal{L}_{\underline b}} \boldsymbol{f}_{l,t,s}
= D_{b,t,s}
&& \forall b,t,s
\label{eq:power_balance_sto_tot}
\\
& 0 \le \boldsymbol{p}_{g,t,s} \le \bar P_g
&& \forall g,t,s
\label{eq:gen_capacity_sto_tot}
\\
& -\bar F_l
\le \boldsymbol{f}_{l,t,s} \le
\bar F_l
&& \forall l \in \mathcal{L},t,s
\label{eq:flow_limit_sto_ms_tot}
\\
& -\bar F_l \boldsymbol{y}_{l,s}^{\text{built}}
\le \boldsymbol{f}_{l,t,s} \le
\bar F_l y_{l,s}^{\text{built}}
&& \forall l \in \mathcal{L}^C,t,s
\label{eq:investment_limit_sto_ms_tot}
\\
& \sum_{b,t} \boldsymbol{LS}_{b,t,s}
\le
\gamma \sum_{b,t} D_{b,t,s}
&& \forall s
\label{eq:LS_req_sto_ms}
\\
& \boldsymbol{LS}_{b,t,s} \ge 0
&& \forall b,t,s
\label{eq:LS_pos_sto_ms}
\\
& \boldsymbol{y}_{l,s}^{\text{inv}},\;
\boldsymbol{y}_{l,s}^{\text{built}},\;
\tilde{\boldsymbol{y}}_{l,s} \in \{0,1\}
&& \forall l \in \mathcal{L}^C, s
\label{eq:binary_investment_sto_ms_tot}
\\
& \boldsymbol{y}_{l,s'}^{\text{built}} = \boldsymbol{y}_{l,s}^{\text{inv}},\;
\boldsymbol{y}_{l,s'}^{\text{built}} \ge \boldsymbol{y}_{l,s}^{\text{built}}
&& \forall l \in \mathcal{L}^C,(s,s')\in\mathcal{P}
\label{eq:lead_time_consistency_tot}
\\
& \boldsymbol{y}_{l,s}^{\text{inv}} - \boldsymbol{y}_{l,s}^{\text{built}}
= \tilde{\boldsymbol{y}}_{l,s}
&& \forall l \in \mathcal{L}^C,s
\label{eq:unique_inv_tot}
\end{align}

The objective function \eqref{eq:objective_sto_ms_disc_tot} minimises the expected total cost across all scenarios, including annualised investment and discounted operational costs. Constraint \eqref{eq:power_balance_sto_tot} enforces nodal power balance for each bus, time step, and scenario. Generator outputs are limited by their capacities in \eqref{eq:gen_capacity_sto_tot}. Transmission line flows are bounded by line capacities for existing lines \eqref{eq:flow_limit_sto_ms_tot} and by investment-dependent capacities for candidate lines \eqref{eq:investment_limit_sto_ms_tot}. Constraint \eqref{eq:LS_req_sto_ms} ensures compliance with a given reliability standard, while \eqref{eq:LS_pos_sto_ms} enforces non-negativity.


Investment decisions are represented by binary variables indicating investment, operational status, and newly built lines \eqref{eq:binary_investment_sto_ms_tot}. Lead time constraints enforce that lines become operational one stage after investment using the transition set $\mathcal{P}$. Eq. \eqref{eq:lead_time_consistency_tot} links investment and operation decisions, ensuring persistence of built lines across stages. Eq. \eqref{eq:unique_inv_tot} links the binary variables to ensure that investment costs are counted only once.

\section{Surrogate model training for learning operational decisions in transmission planning}
\label{sec:LOD}

\begin{figure}
    \centering
    \includegraphics[width=\linewidth]{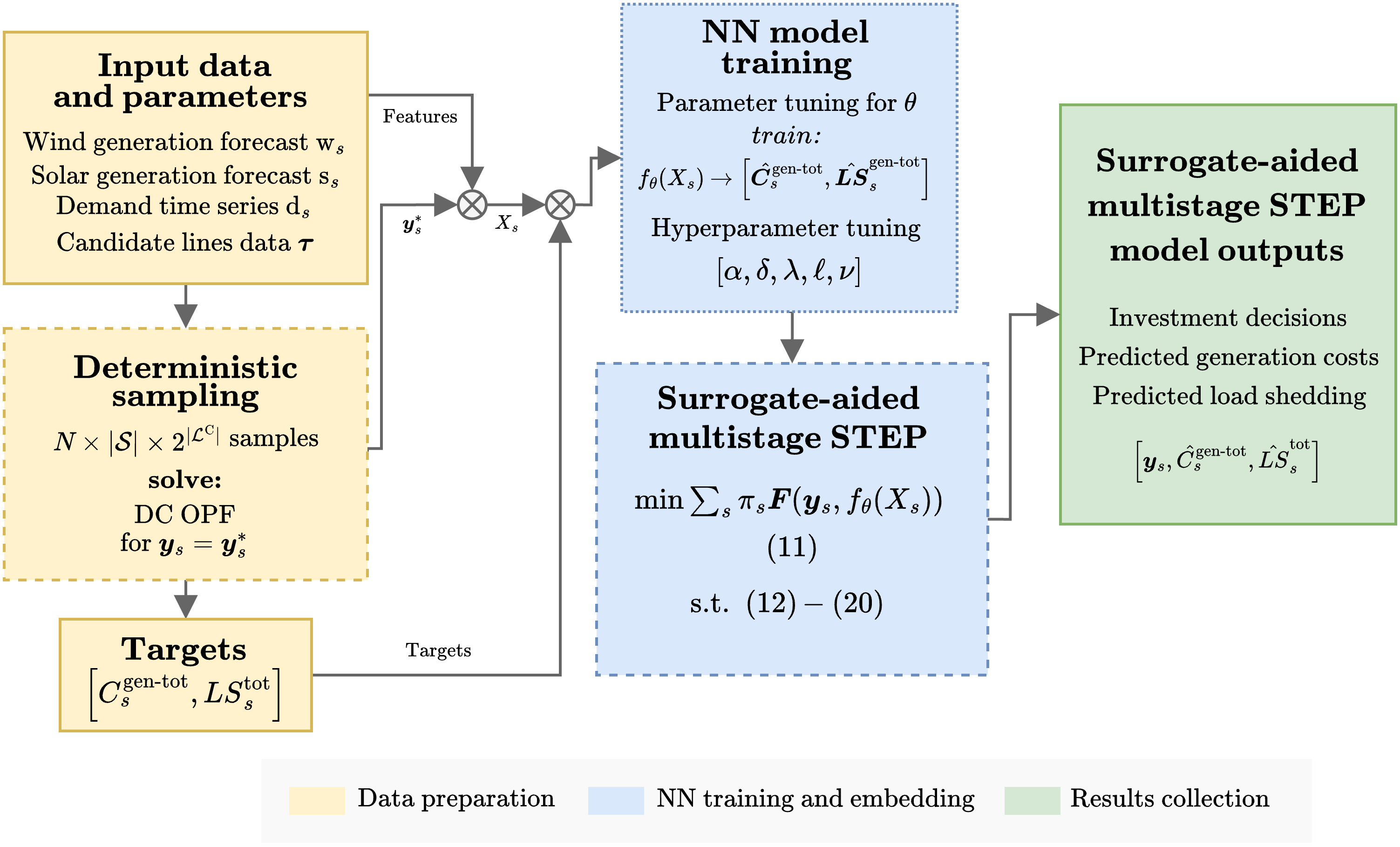}
    \caption{ML embedding framework. Yellow boxes denote the data preparation steps. Blue indicates the NN preparation and embedding. Green represents the outputs obtained, which are benchmarked against the results of the multistage exact formulation.}
    \label{fig:ML_pipeline}
\end{figure}

Expanding on the approach proposed in \cite{dumouchelle2022neur2spneuraltwostagestochastic}, this work proposes deriving the operational decisions of each node $s$ of the STEP model, as represented in Section~\ref{sec:model_sto} with a NN trained to predict annual operational decisions for each node $s$. Total generation costs $C^{\text{gen-tot}}_{s}$ and total load shedding $LS^{\text{tot}}_{s}$ are approximated by a parametric model $f_{\theta}(\cdot)$ with parameters (weights and biases) $\theta$. The input vector is defined as $X_{s} = [\text{w}_s,\; \text{s}_s,\; \text{d}_s,\; \boldsymbol{\tau},\: \boldsymbol{y}_s]$, which combines scenario-dependent parameters and investment decision variables $\boldsymbol{y}_s$. The surrogate model evaluates $\small f_{\theta}(X_s(\boldsymbol{y}_s))  \rightarrow \left[\hat{\boldsymbol{C}}^{\text{gen-tot}}_{s}, \hat{\boldsymbol{LS}}^{\text{tot}}_{s}\right]$, making the predicted operational values explicitly dependent on the investment decisions, while the hourly operational constraints are implicitly captured in the neural network parameters. The overall workflow presenting this approach is illustrated in Fig.~\ref{fig:ML_pipeline}.

\subsection{Data Generation and Sampling}
\label{ref:data_gen}
A deterministic sampling procedure is used to generate the training dataset $X_s$. This process can be interpreted as a scenario-based decomposition. Let $N$ denote the number of distinct training runs. For each run, the operational problem is solved for every scenario $s \in \mathcal{S}$ and each possible network configuration. Each configuration is represented by a unique binary vector of candidate line investments, resulting in $2^{|\mathcal{L}^{\text{C}}|}$ combinations, with $\mathcal{L}^{\text{C}}$ denoting the set of candidate lines. To improve generalisation and explicitly capture uncertainty, random scaling factors $\kappa \sim U([-0.25,0.25])$ are applied $N$ times for each node $s$ to wind ($\text{w}_{s}$), solar ($\text{s}_{s}$), and demand ($\text{d}_{s}$) profiles around baseline values. Each profile for individual generators and loads is perturbed independently at each node $s$, avoiding induced correlations while exposing the trained surrogate model to a broader range of plausible operating conditions.


The total number of deterministic problems solved to generate the training dataset is $N \times |\mathcal{S}| \times 2^{|\mathcal{L}^{\text{C}}|}$. $N$ is chosen according to the complexity of the system and the NN architecture required for satisfactory predictive accuracy. Each run solves an optimal power flow (OPF) for a given network configuration $\Gamma$ and scenario node $s$. The solution of each OPF produces a row of the input features $X_{s}$ and the target outputs, namely, the yearly total generation cost $(C^{\text{gen-tot}}_s)$ and the total load shedding $(LS^{\text{tot}}_{s})$. This procedure ensures that the training dataset fully captures the relationship among decisions on transmission investments, scenario realisations, and resulting operational decisions.

\subsection{Network Design and Architecture}
To predict total operational costs and load shedding, a feed-forward, fully connected NN is used. For a NN with $\ell$ hidden layers, the mapping $f_{\theta}(X_s)$ can be expressed as:
\begin{align}
    h_1 =\ & \sigma(W_0 X_s + b_0), \notag \\
    h_{m+1} =\ & \sigma(W_m h_m + b_m), \quad m = 1,\dots,\ell-1, \notag \\
    [\hat{\boldsymbol{C}}^{\text{gen-tot}}_{s}, \hat{\boldsymbol{LS}}^{\text{tot}}_{s}] =\ & W_\ell h_\ell + b_\ell, \notag
\end{align}
$h_m$ denotes the activation of layer $m$. $W_m$ and $b_m$ are the weights and biases, while $\sigma(\cdot)$ is the activation function (here ReLU). The \textit{Huber} loss function is used for training due to its desirable properties: near zero, it behaves like the squared loss, ensuring smooth gradients and stable convergence, while for large residuals, it transitions to a linear loss, reducing sensitivity to outliers \cite{hastie2009elements}. Models are trained using the \textit{Adam} optimiser with batch sizes of 64 (for total generation cost) and 32 (for total load shedding) for up to 5000 epochs with early stopping on validation loss (patience 250) and a \textit{ReduceLROnPlateau} scheduler reducing the learning rate by 25\% after 75 stagnant epochs, bounded below by $10^{-6}$.

\subsection{Hyperparameter Optimisation}
To ensure accurate predictions, the NN hyperparameters are optimised separately for each target variable. For each target, a Bayesian search is performed over predefined ranges of candidate values to identify the combination that minimises the validation loss. Table~\ref{tab:hyperparameters_choice} summarises the search ranges considered for each hyperparameter.

\begin{table}
\centering
\caption{Search ranges for hyperparameters}
\begin{tabular}{ll}
\toprule
\textbf{Hyperparameter} & \textbf{Search Range}\\
\midrule
Initial learning rate $(\alpha)$ & $10^{-5}$ to $10^{-2}$ \\
Huber loss delta $(\delta)$ & $10^{-4}$ to $10$ \\
L2 kernel regularisation coefficient $(\lambda)$ & $10^{-6}$ to $10^{-2}$ \\
Number of hidden layers $(\ell)$ & $2$ to $3$ \\
Units per hidden layer $(\nu)$ & $32$ to $512$ (in steps of 32) \\
\bottomrule
\end{tabular}
\label{tab:hyperparameters_choice}
\end{table}
 
The model configuration that achieves the lowest validation loss is chosen as the best performer. To minimise the risk of overfitting to the validation set, the final model is re-evaluated on a separate testing set. This provides an unbiased estimate of the model's generalisation performance, ensuring that the reported predictive accuracy reflects the model's robustness, rather than being influenced by bias introduced during hyperparameter tuning.

\subsection{Optimisation model ReLU embedding}
Once trained, the NN model $f_{\theta}(\cdot)$ can be embedded into the stochastic optimisation model to approximate operational decisions. Following the formulations proposed in \cite{dumouchelle2022neur2spneuraltwostagestochastic, Fischetti2018}, embedding a trained NN into a mixed-integer linear program (MILP) requires linearising the ReLU activation function. This reformulation allows the NN to be incorporated as linear constraints into the optimisation model. ReLU networks are used here, as they consist of max-affine operators that admit an exact MILP formulation, whereas alternative nonlinear or smooth convex surrogates (e.g., sigmoid or quadratic models) require additional approximations, resulting in larger and less efficient optimisation formulations \cite{Grimstad_2019}. The original STEP formulation \eqref{eq:objective_sto_ms_disc_tot}–\eqref{eq:unique_inv_tot} can then be written as:


\begin{align}
    \min_{\substack{\tilde{\boldsymbol{y}}_{l,s}, \boldsymbol{y}^{\text{built}}_{l,s}, \boldsymbol{y}^{\text{inv}}_{l,s}, \\ \hat{\boldsymbol{C}}^{\text{gen-tot}}_s, \hat{\boldsymbol{LS}}^{\text{tot}}_s }}
    & \sum_{s \in \mathcal{S}} \sum_{l \in \mathcal{L}^C} \pi_s \cdot c_{l,s}^{\text{ann}} \cdot \tilde{\boldsymbol{y}}_{l,s} + \notag\\ 
    & \sum_{s \in \mathcal{S}} \frac{\pi_s}{(1 + r)^{\text{year}_s}} \left(
    \hat{\boldsymbol{C}}^{\text{gen-tot}}_s + 
    \text{VoLL} \cdot \hat{\boldsymbol{LS}}^{\text{tot}}_s \right)
    \label{eq:objective_embedding} 
\end{align}

\begin{align}
\text{s.t.}\quad
&\sum_{i=1}^{|\mathcal{D}_0|} w_{ij}^0 [\mathbf{y}^{\text{built}}_s, X_s]_i + b_j^0 
= \overline{h}_j^{1,s} - \underline{h}_j^{1,s} \quad \forall j \in \mathcal{D}_1, s 
\label{eq:first_layer_embedding} \\
&\sum_{i=1}^{|\mathcal{D}_{m-1}|} w_{ij}^{m-1} \overline{h}_i^{m-1,s} + b_j^{m-1} 
= \overline{h}_j^{m,s} - \underline{h}_j^{m,s}, \notag \\
& \forall m \in [1,\ell-1], j \in \mathcal{D}_m, s
\label{eq:hidden_layers_embedding} \\
&\sum_{i=1}^{|\mathcal{D}_{\ell}|} w_{ik}^{\ell} \overline{h}_i^{\ell,s} + b^{\ell}
\leq \{\hat{\boldsymbol{C}}^{\text{gen,tot}}_s,\hat{\boldsymbol{LS}}^{\text{tot}}_s\}_k, \quad \forall k\in\{1,2\}, s 
\label{eq:output_layer_embedding} \\
&\overline{h}_j^{m,s} \le M(1-z_j^{m,s}), \;
\underline{h}_j^{m,s} \le M z_j^{m,s}, \notag \\
& \forall m \in [1,\ell-1], j \in \mathcal{D}_m, s 
\label{eq:relu_bigM} \\
&z_j^{m,s}\in\{0,1\}, \;
\overline{h}_j^{m,s},\underline{h}_j^{m,s}\ge0, \notag \\
& \forall m \in [1,\ell-1], j \in \mathcal{D}_m, s 
\label{eq:relu_domains} \\
&\hat{\boldsymbol{C}}^{\text{gen,tot}}_s,\hat{\boldsymbol{LS}}^{\text{tot}}_s \ge 0, \quad \forall s 
\label{eq:positive_output} \\
%
&\eqref{eq:binary_investment_sto_ms_tot} - \eqref{eq:unique_inv_tot} \notag
\end{align}

The objective \eqref{eq:objective_embedding} minimises investment costs and predicted expected operational costs. NN propagation is embedded explicitly: the input layer feeds the first hidden layer as defined in \eqref{eq:first_layer_embedding}, and subsequent hidden layers propagate activations according to \eqref{eq:hidden_layers_embedding}. The output layer \eqref{eq:output_layer_embedding} produces predictions of total generation cost and load shedding for each scenario, represented by $\hat{\boldsymbol{C}}^{\text{gen-tot}}_s$ and $\hat{\boldsymbol{LS}}^{\text{tot}}_s$.

ReLU activation functions are modelled using a big-$M$ formulation with binary activation variables $z_j^{m,s}$ and domain constraints \eqref{eq:relu_bigM}-\eqref{eq:relu_domains}. Positivity of the predicted outputs is enforced in \eqref{eq:positive_output}. Each NN component is scenario-indexed to capture scenario-dependent operational outcomes. Eqs. \eqref{eq:binary_investment_sto_ms_tot}–\eqref{eq:unique_inv_tot} are retained from the previous formulation to ensure the temporal consistency of transmission investments.

\section{Numerical studies}
This section presents the results of case studies conducted on the IEEE 33-bus and IEEE RTS-24 test systems. All simulations were conducted in \textit{Python}, with optimisation problems formulated and solved using \textit{Gurobi v12.0.1}, and ML models implemented using \textit{TensorFlow}. The optimisation problems were solved with a time limit of 1,000 seconds, a relative MIPgap of 0.1\%, and default feasibility and optimality tolerances of $10^{-6}$. Computations were performed on a University of Melbourne workstation featuring an AMD EPYC 9474F processor (3.6 GHz) with 48 physical cores and 96 threads, and 128 GB of RAM. 

\subsection{Case study \#1: IEEE 33-bus system}
\begin{figure}
    \centering
    \includegraphics[width=\linewidth]{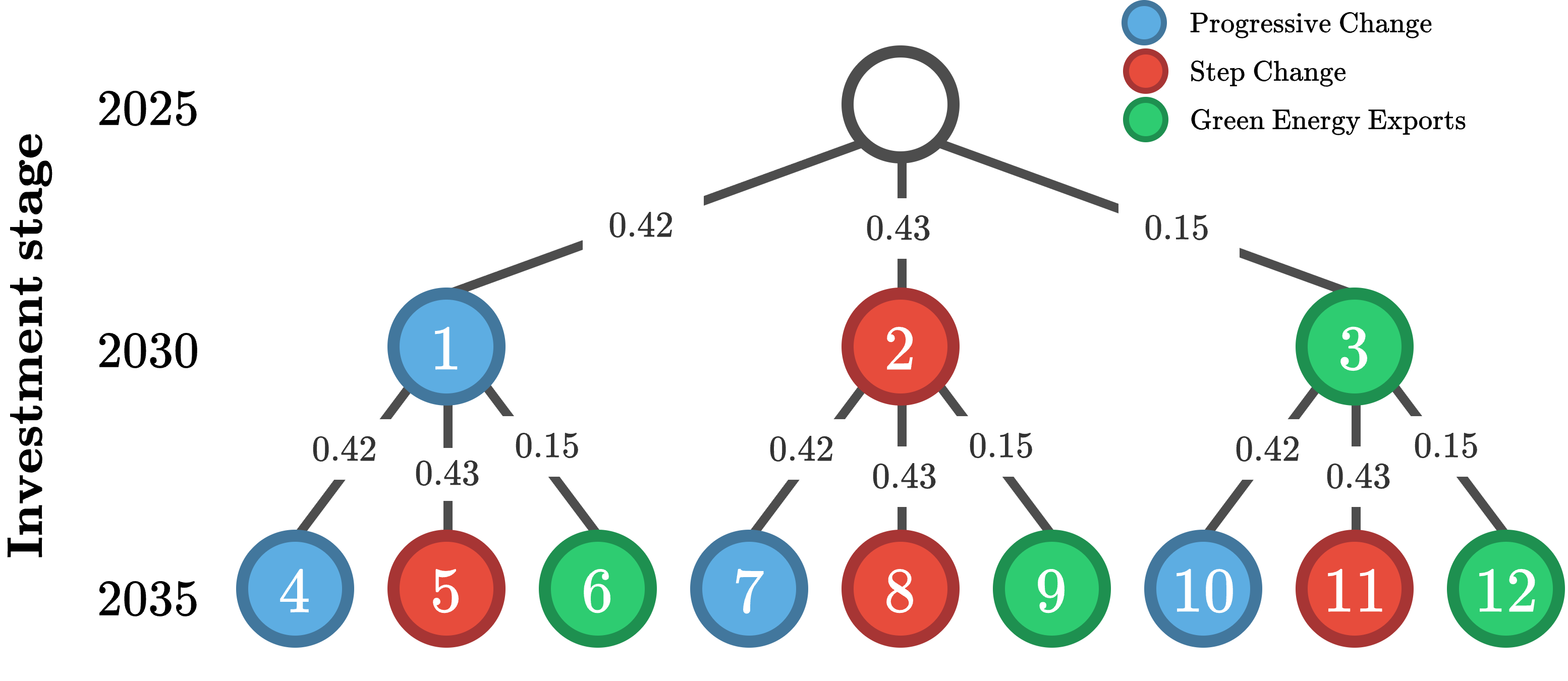}
    \caption[Scenario tree considered.]{Scenario tree considered. A 15-year decision horizon is considered. Each colour refers to a development path based on Australia's east coast power system 2024 Integrated System Plan \cite{2024_ISP}.}
    \label{fig:scenario_tree_isp}
\end{figure}

The first system studied is the IEEE 33-bus system. The candidate lines considered here are lines noted as \textit{switchable}. The technical parameters of the lines and generation limits, as well as the maximum load supported by the system in the base case (\textit{i.e.} year 2025, the root node of the scenario tree considered, shown on Fig. \ref{fig:scenario_tree_isp}) are taken from \cite{IEEE_33_bus}. A Value of Lost Load (VoLL) of \$15,000 AUD/MWh is set to penalise load shedding. 

\subsubsection{Input Data}
\label{ref:data_IEEE33}
The active power demand over a 24-hour horizon is taken from \cite{IEEE_33_bus} and represents the hourly aggregated demand across the entire system. This demand profile serves as the daily base load for the root node of the scenario tree by setting the maximum demand for that year (\textit{i.e.} 2025).\\
Generation capacity data, adapted from \cite{IEEE_33_bus}, has been scaled and adjusted to ensure feasibility with respect to the load-shedding reliability requirement during data sampling. This modification is necessary to enable evaluation of scenarios in which insufficient generation results in unserved energy, thereby activating/violating the model's reliability constraints. Indeed, load shedding must comply with the Reliability Standard, which requires that at least ($100-\gamma$)\% of demanded energy is met for a given financial year. This requirement is defined, for example, under \textit{Clause 3.9.3C} of the Australian National Electricity Rules \cite{NER_3.9.3C}, and is usually set as $\gamma = 0.002$\%.

\begin{table}
    \centering
    \caption{Candidate line data: Capacity and investment costs  - IEEE 33-bus study case.}
    \renewcommand{\arraystretch}{1} 
    \setlength{\tabcolsep}{7pt} 
    \begin{tabular}{ccccc}
        \toprule
        Line \# & From & To & Capacity [MW] & Investment\\
        & Node & Node & & Cost [k\$/MW] \\ \midrule
        33 & 8 & 21 & 1.0 & 100 \\
        34 & 12 & 22 & 1.0 & 100 \\
        35 & 25 & 29 & 3.8 & 100 \\
        \bottomrule
    \end{tabular}
    \label{tab:candidate_lines_IEEE33}
\end{table} 

\subsubsection{Framework application}
Following the target data generation procedure described in \ref{ref:data_gen}, a comprehensive training dataset of input features and targets is obtained after $N=500$ training runs. $N$ is defined via a sensitivity analysis to ensure marginal improvements beyond this point. The NN training framework is then applied with and without hyperparameter optimisation (HPO). In that case, generating the 500 samples requires 9 hours, training each model takes 1 hour and 40 minutes, and HPO lasts 10 hours per model. Note that 20\% of the generated dataset is allocated to the validation set. To address data sparsity in load shedding events, a stratified splitting strategy is applied to ensure that scenarios with and without load shedding are proportionally represented in both the training and validation sets. Training and validation loss curves are monitored to assess convergence and detect overfitting. Comparative training curves for models with and without HPO reveal clear trade-offs, highlighting differences in performance and the gains achieved through hyperparameter optimisation. 

\begin{figure}
    \centering
    \begin{subfigure}[c]{\linewidth}
        \centering
        \includegraphics[width=\linewidth]{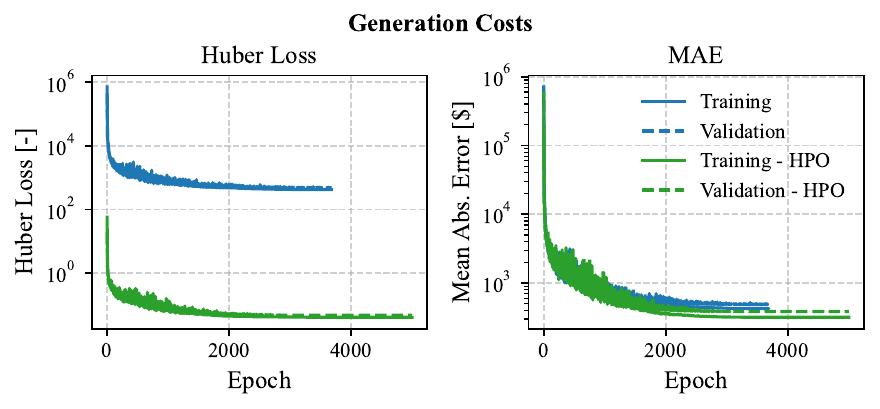}
        \caption{Training and validation loss for generation costs.}
        \label{fig:training_HPO_gen_costs}
    \end{subfigure}
    \vfill
    \begin{subfigure}[c]{\linewidth}
        \centering
        \includegraphics[width=\linewidth]{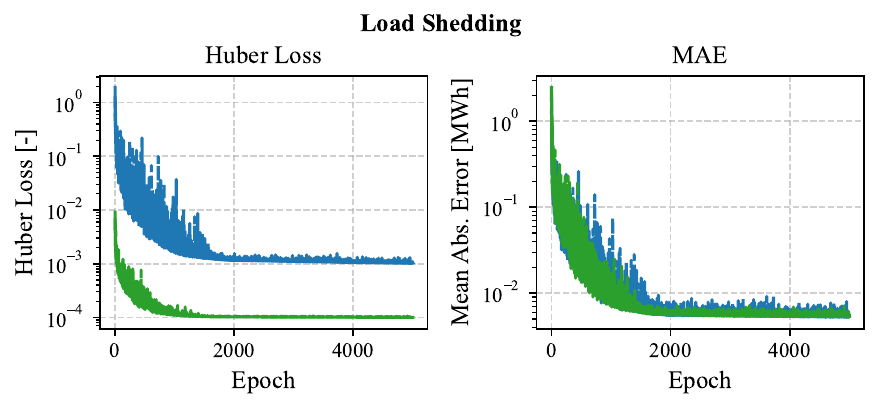}
        \caption{Training and validation loss for Load Shedding.}
        \label{fig:training_HPO_LS}
    \end{subfigure}
    \caption{Training and validation loss curves with and without hyperparameter optimisation.}
    \label{fig:training_HPO}
\end{figure}



\subsubsection{Trustworthiness}
\begin{figure}
    \centering
    \begin{subfigure}[c]{0.49\textwidth}
        \centering
        \includegraphics[width=\linewidth]{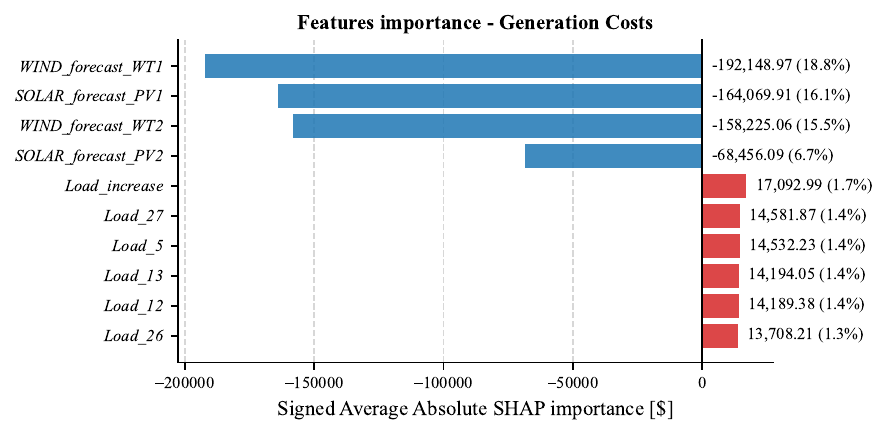}
        \caption{SHAP importance for generation costs.}
        \label{fig:shap_gen_costs}
    \end{subfigure}
    \vfill
    \begin{subfigure}[c]{0.49\textwidth}
        \centering
        \includegraphics[width=\linewidth]{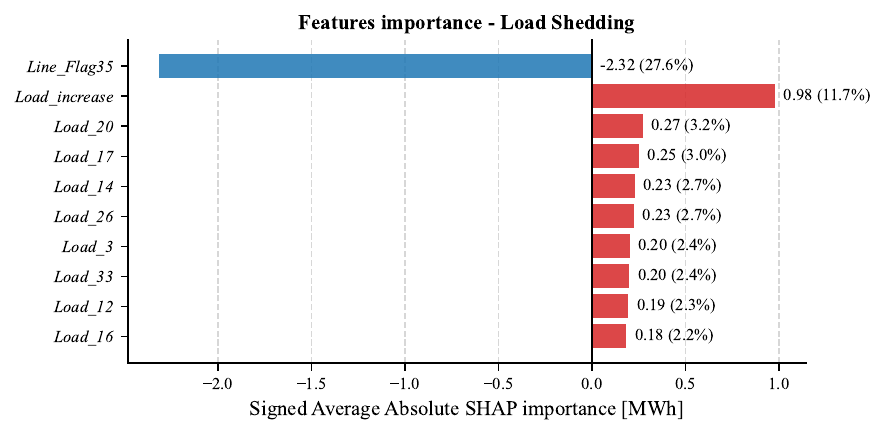}
        \caption{SHAP importance for load shedding.}
        \label{fig:shap_LS}
    \end{subfigure}
    \caption{Signed SHAP importance for the ten most influential features in the surrogate model. Bar length represents the mean absolute SHAP value (magnitude of influence), while bar direction indicates the sign of the mean SHAP value (positive or negative contribution to the model output). Percentages denote the relative contribution of each feature.}
    \label{fig:shap_combined}
\end{figure}

As transmission energy systems are part of critical infrastructure, SHAP values were employed in this study to enhance the transparency of the surrogate models. The SHAP explainer \cite{NIPS2017_8a20a862} was given 10,000 training samples as background distribution and was evaluated on an additional 1,000 samples to ensure consistent estimations of feature importance. The background dataset is used to generate samples, which are used to calculate these estimates.\\
Figs.~(\ref{fig:shap_gen_costs}) and (\ref{fig:shap_LS}) present the results for the 1,000 evaluation samples (scatter data), illustrating how the models respond to variations in the input features. Feature values that increase the model output relative to the mean are shown in red, whereas those that decrease it are shown in blue.


The results are consistent with domain knowledge. Generation costs decrease on average as renewable generation increases (negative mean SHAP; mean $\left| \text{SHAP} \right|$ = $192,148.47$\$ for feature \textit{WT1}), while rising load levels exert a positive contribution to costs, reflecting increased generation requirements. For load shedding, investment in Line~35 exhibits a strong mitigating effect, contributing more than 27\% to the reduction of expected load shedding due to the associated increase in network transfer capacity. In contrast, higher load levels significantly increase both the probability and magnitude of load shedding, consistent with system operation near capacity limits. Overall, the top 10 features account for 65.7\% of the total mean absolute SHAP importance for Generation Costs and 60.2\% for Load Shedding, confirming that the model’s behaviour is primarily driven by physically meaningful variables. This analysis therefore demonstrates that Shapley values provide both global interpretability (through feature ranking based on mean $\left| \text{SHAP} \right|$) and local interpretability (through instance-level decomposition), thereby strengthening the transparency and trustworthiness of the proposed ML pipeline.

\subsubsection{Testing Results}
\label{sec:IEEE_33_res}

To evaluate the final model, 5 testing instances were generated from the reference load and generation profiles in \ref{ref:data_IEEE33}. Unseen noise is added using the same uniform distribution ($\kappa \sim \mathcal{U}([-0.25, 0.25])$) as in training. Only five instances were considered as solving additional full stochastic optimisation benchmark results is computationally expensive. Further instances would primarily confirm the observed trend in computational benefits. For this evaluation, the parameters of the candidate transmission lines are fixed to their true values, as listed in Table~\ref{tab:candidate_lines_IEEE33}.

\begin{figure}
    \centering
    \begin{subfigure}[c]{0.49\linewidth}
        \centering
        \includegraphics[width=\linewidth]{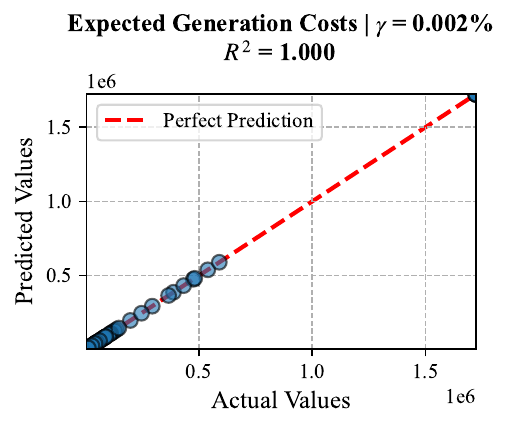}
        \caption{Prediction accuracy for generation costs.}
        \label{fig:regression_gen_costs}
    \end{subfigure}
    \hfill
    \begin{subfigure}[c]{0.48\linewidth}
        \centering
        \includegraphics[width=\linewidth]{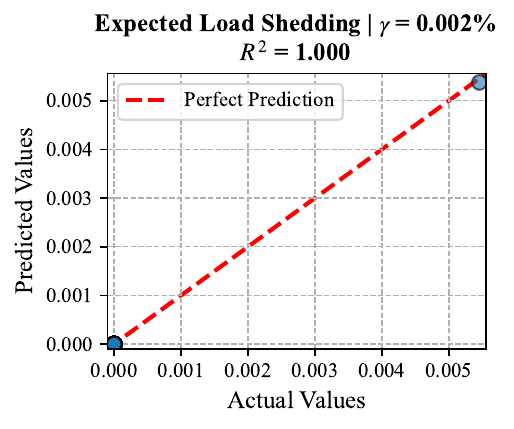}
        \caption{Prediction accuracy for load shedding.}
        \label{fig:regression_ls}
    \end{subfigure}
    \caption{Regression performance as a function of training set size for (a) generation costs and (b) load shedding. Note that for Load Shedding, most of the points are located at extremities of the dashed line.}
    \label{fig:regression_performance}
\end{figure}

The predictive performance of the surrogate model is assessed using the Mean Absolute Error (MAE) and the $R^2$ score. Figs.~(\ref{fig:regression_gen_costs}) and~(\ref{fig:regression_ls}) present the prediction accuracy of these surrogate models trained to approximate the expected generation costs and load shedding, respectively. Both plots show the predicted versus actual values across the test dataset, alongside the reference line indicating perfect prediction, taken from the multistage exact formulation.

\begin{figure}
    \centering
    \includegraphics[height=1\linewidth, angle=-90]{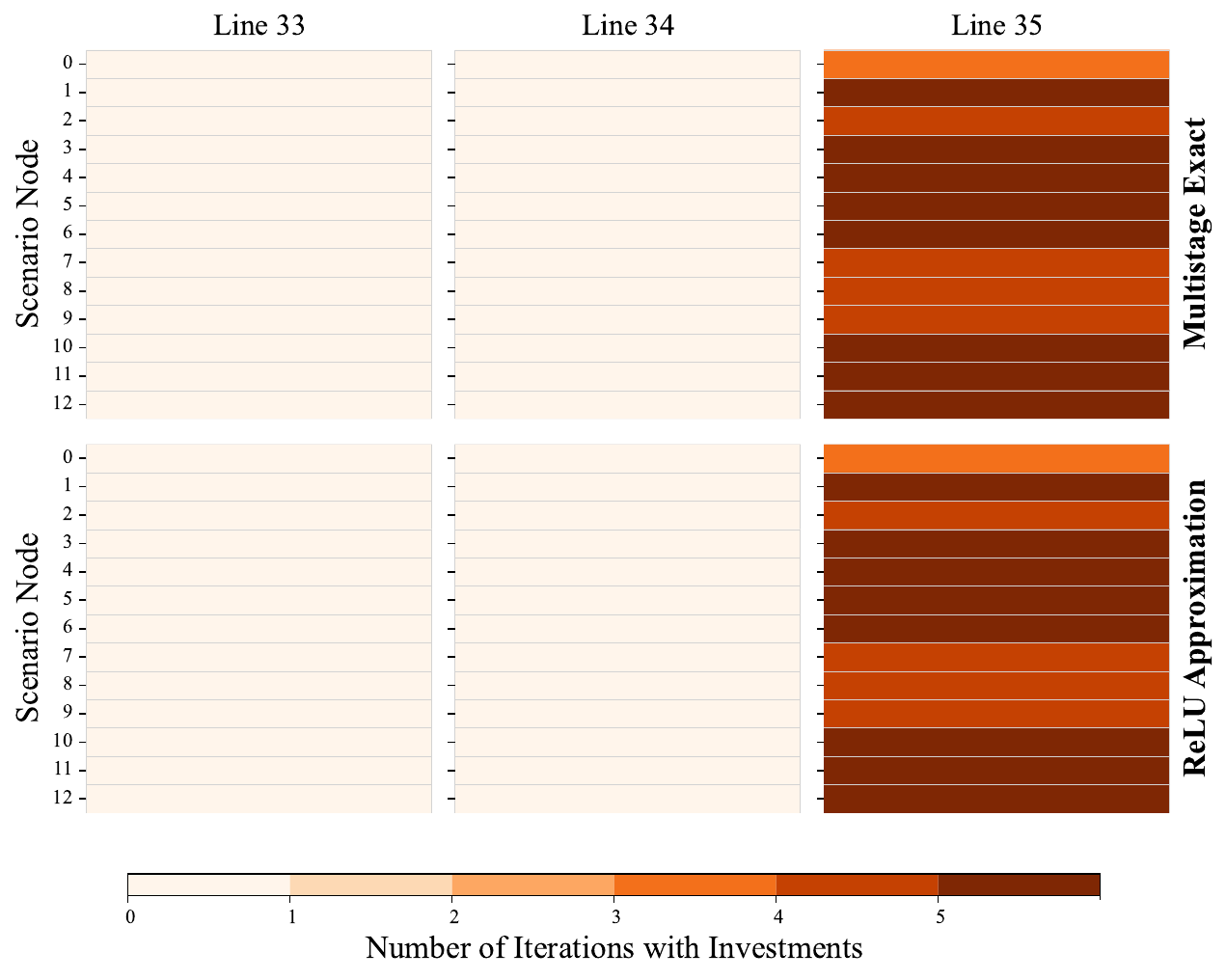}
    \caption[Investment decisions for the line candidates.]{Investment decisions for the line candidates. The heatmap displays the total number of investments for each line candidate, per scenario node across all 5 testing instances.}
    \label{fig:investment_decisions}
\end{figure}

In both cases, the results demonstrate an excellent fit, with a $R^2$ score of 1.000 and negligible deviation from the ideal line. This confirms that the surrogate models accurately capture the input–output mapping and provide reliable approximations when embedded within the stochastic optimisation framework.
Fig.~\ref{fig:investment_decisions} shows the predicted investment decisions for the three line candidates across different testing instances. The plot compares the binary investment outcomes obtained using the exact multistage stochastic formulation with those produced by the surrogate-based ReLU approximation. For Lines~33 and~34, no investments were undertaken in either the exact or approximate models. For Line~35, the surrogate decisions closely follow the exact solutions. 

This indicates that the surrogate model reliably captures the key structural drivers influencing whether these lines should be upgraded or not under varying operational conditions. Note that the number of investments per scenario node depends on the noise distribution adopted for each testing instance.

\begin{figure}
    \centering    
    \includegraphics[width=\linewidth]{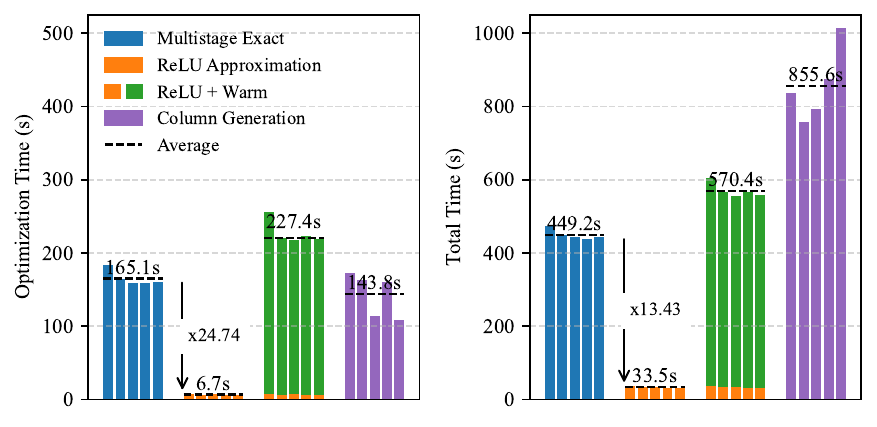}
    \caption{Solving and Total run time comparison.}
    \label{fig:time_comparison_warm}
\end{figure}

Fig.~\ref{fig:time_comparison_warm} compares the computational performance of the exact multistage stochastic formulation with the surrogate-based ReLU approximation, considering both solver time and total execution time per testing instance. The results demonstrate the practical benefits of surrogate models \textit{after training}, achieving up to a 13.4$\times$ reduction in solution time while maintaining decision quality and feasibility. Note that the training phase takes around 30 hours, but once completed, the surrogate model can quickly evaluate many grid configurations and scenarios without repeatedly solving the full operational problem. Fig.~\ref{fig:time_comparison_warm} also reports the solving and total time when benchmarking the proposed approach against CG, a commonly adopted decomposition technique for STEP problems, showing a speed-up factor of approximately 20.

This computational improvement is largely driven by the reduction in optimisation model size achieved with the surrogate formulation. In the explicit stochastic operational formulation, the operational problem leads to a very large model with 8,313,357 variables (8,313,240 continuous and 117 binary) and 12,299,158 constraints, representing decisions such as generation dispatch, power flows, and load shedding across all time steps and scenarios. In contrast, replacing the operational problem with the NN surrogate embedded in the planning model results in a MILP with only 35,087 variables (23,322 continuous and 11,765 binary) and 35,088 constraints. This reduction of roughly two orders of magnitude in model size significantly lowers the computational burden and explains the observed runtime improvements. Consequently, the surrogate-based formulation can be efficiently executed on standard personal computers without requiring high-performance computing resources.

Moreover, one approach to guarantee the optimality of the surrogate model’s investment decisions is to pass them as initial solutions to the multistage exact formulation in order to test them. Fig.~\ref{fig:time_comparison_warm} reports also the resulting solving times when the solver is warm-started with the surrogate’s suggested investment plan. However, the results show that warm-starting the solver with surrogate outcomes does not actually reduce overall solving time. In contrast, it slightly increases computation time, even though the initial investment scheme is both feasible and optimal. This suggests that when the surrogate’s solution is already very close to optimality, the overhead of warm-start processing may outweigh its potential benefits.

\subsubsection{Sensitivity Analysis on the Noise Range} A sensitivity analysis is conducted to evaluate the model’s robustness under different noise levels. One important case examined is the one in which no perturbation is made. This corresponds to solving the original stochastic problem and is particularly relevant in practice, since state-of-the-art solvers may fail to find optimal solutions due to problem size and complexity. In such a case, ML-enhanced optimisation can provide feasible solutions to otherwise intractable problems. The analysis considers noise levels ranging from 0\% (the unperturbed stochastic problem) up to 125\% of the original perturbation range used in training ($\kappa = 31.25\%$). This setup tests the model’s reliability under both ideal and increasingly noisier conditions.

\begin{figure}
    \centering    
    \includegraphics[width=\linewidth]{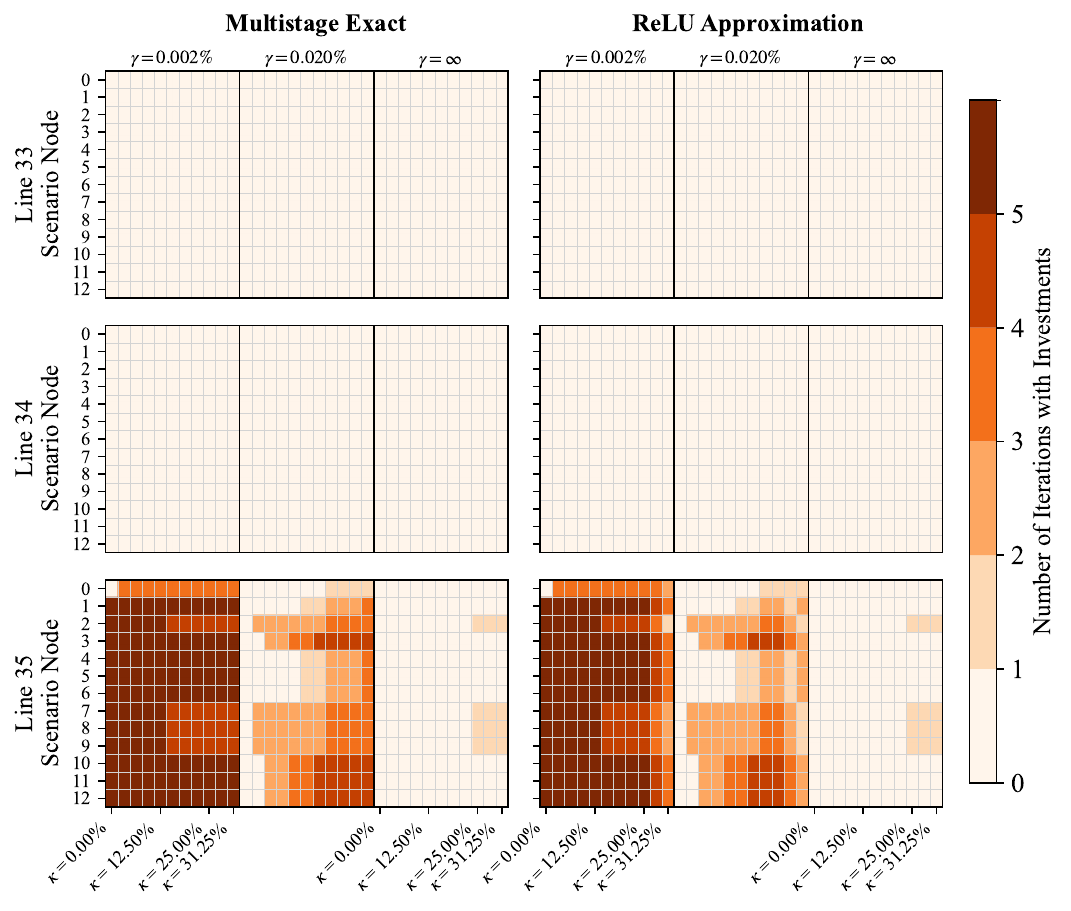}
    \caption{Investment decisions comparison across different $\gamma$ and $\kappa$ settings.}
    \label{fig:inv_delta}
\end{figure}

Fig.~\ref{fig:inv_delta} compares the investment decisions obtained from the multistage exact model and the ReLU-based approximation across different levels of perturbation $\kappa$ and reliability requirements $\gamma$.

The results show a strong overall match, with investment patterns remaining consistent across most scenarios. However, once perturbations exceed the training range, the surrogate starts to diverge from the exact solution. This highlights its sensitivity to inputs outside the training distribution and suggests a degradation in the reliability of the surrogate approach near or beyond this boundary, consistent with patterns observed in the generation cost and load-shedding error distributions.

\subsection{Case study \#2: IEEE RTS 24-bus system}

To demonstrate the tractability of the proposed methodology across different grid configurations, the framework is also applied to the 24-bus IEEE Reliability Test System (RTS) \cite{ordoudis2016updated}. The system model accommodates six renewable generators, enabling acase study with higher renewable energy penetration.

\subsubsection{Input Data}
The technical data of the lines and generation limits, as well as the maximum load supported by the system in the base case (\textit{i.e.} here 2025, the root node of the scenario tree considered), are taken from \cite{ordoudis2016updated}. Three lines in the system have been designated as candidate lines for expansion. Table~\ref{tab:candidate_lines_IEEE24} summarises the candidate lines available for investment. 

\begin{table}
    \centering
    \caption{Candidate Line Data with Capacity and Investment Costs - IEEE 24-bus Study Case.}
    \renewcommand{\arraystretch}{1}
    \setlength{\tabcolsep}{7pt} 
    \begin{tabular}{ccccc}
        \toprule
        Line \# & From & To & Capacity [MW] & Investment\\
        & Node & Node & & Cost [k\$/MW] \\
        \midrule
        35 & 18 & 21 & 100 & 500 \\
        36 & 15 & 22 & 100 & 500 \\
        37 & 13 & 23 & 100 & 500 \\
        \bottomrule
    \end{tabular}
    \label{tab:candidate_lines_IEEE24}
\end{table}

\subsubsection{Testing Results}
The same sampling and training framework described in Section~\ref{sec:LOD} is applied here. To evaluate the developed model, five test instances are randomly generated following the procedure outlined in Section~\ref{sec:IEEE_33_res}.


When evaluating the surrogate model performance, it again offers a clear computational advantage. Solver times are consistently shorter compared to the exact formulation. Furthermore, when the model encounters infeasible instances, the ReLU surrogate enables a quick identification. 
\begin{figure}
    \centering    
    \includegraphics[width=\linewidth]{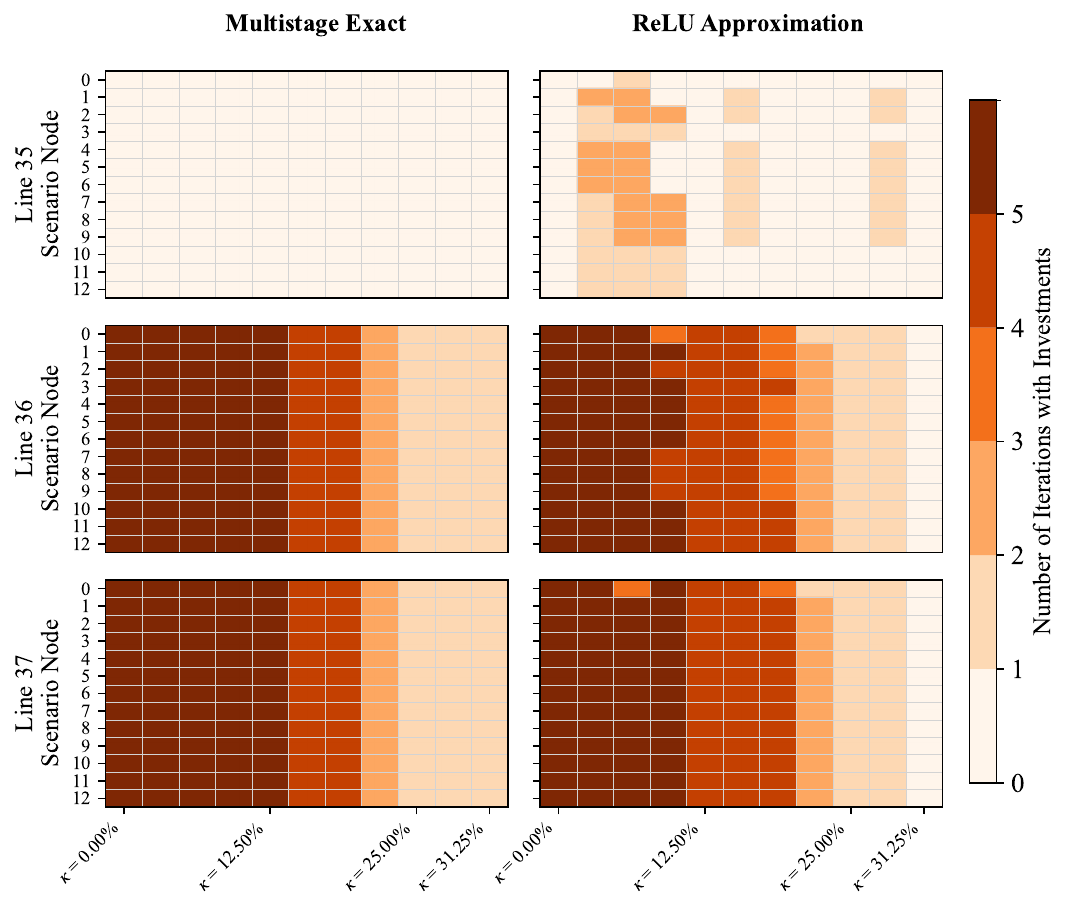}
    \caption{Investment decisions comparison across different $\kappa$ settings.}
    \label{fig:inv_delta_IEEE_24}
\end{figure}

Fig.~\ref{fig:inv_delta_IEEE_24} compares the investment decisions obtained from both the multistage exact model and the ReLU-based surrogate under varying $\kappa$ levels. These heatmaps reveal interesting behaviour: At $\kappa = 0\%$, corresponding to the original multistage problem, the surrogate correctly replicates the investment decisions. As $\kappa$ increases, mismatches begin to appear early, although for some values, the surrogate still achieves perfect alignment, such as for $\kappa = 25\%$. However, as the perturbation level increases, infeasibility becomes more frequent, making direct comparison increasingly difficult.

Overall, while this test system application shows the adaptability of the proposed framework, it also shows an important insight: its performance is closely tied to the quality and structure of the training data. When the underlying patterns are complex or the target variables are relatively sparse, as in the case of Load Shedding, the surrogate model’s accuracy naturally decreases, highlighting opportunities for future improvement through enhanced data generation or tailored NN architectures.

\section{Conclusion}

This work presents a systematic surrogate-based framework for multistage STEP. The framework combines structured data generation with NNs, which are reformulated as sets of linear constraints and integrated within a STEP model to approximate the results of operational decisions. Once the NNs are trained, the approach demonstrates a significant reduction in the computational burden of multistage STEP models while maintaining a high level of predictive accuracy. A key technical contribution of this work lies in enabling fast and reliable surrogate embeddings, thereby making large-scale stochastic planning problems tractable.

Case studies on the IEEE 33-bus and RTS-24 bus systems demonstrate that the proposed framework has potential for scalability, as it enables substantial reductions in solution time while maintaining the fidelity of investment decisions. Importantly, the main value of the approach lies in its ability to repeatedly and efficiently sweep across the space of uncertainty inputs. Once the sampling and training steps, which are performed as one-off tasks, are completed, the trained model can be reused to conduct extensive sensitivity analyses, stress testing, and scenario exploration over full-year chronologies rather than relying on a limited set of representative periods. This enables planners to, for example, assess the robustness of investment plans across the entire probability space without repeatedly solving full stochastic optimisation problems, offering substantial computational advantages and a practical alternative to classical decomposition techniques, which can be cumbersome to implement.

To reach real-world applications, several directions for future research naturally arise. From a methodological perspective, improving guarantees on decision quality when relying on surrogate predictions remains critical, particularly for critical decisions such as investments. Distributionally robust regression methods may offer a promising pathway to formalise such guarantees under uncertainty. From a computational standpoint, scalability could be further enhanced through informed sampling strategies that filter infeasible, redundant, or low-impact line configurations, thereby enabling the consideration of larger networks and richer sets of candidate transmission lines without an exponential increase in data generation costs. In addition, transfer learning techniques could improve model portability across networks, limiting the need to retrain a new model from scratch for each system. Finally, extending the framework to incorporate dynamic grid conditions and the integration of real operational data, including unit commitment, energy storage systems, and more detailed OPF formulations, would further strengthen its applicability to real-world, low-carbon power system planning.

\bibliographystyle{IEEEtran}
\bibliography{ref}

@ARTICLE{Borozan_2025,
  author={Borozan, Stefan and Strbac, Goran},
  journal={IEEE Transactions on Sustainable Energy}, 
  title={{Multi-Stage Integrated Transmission and Distribution Expansion Planning Under Uncertainties With Smart Investment Options}}, 
  year={2025},
  volume={16},
  number={1},
  pages={546-559},
  doi={10.1109/TSTE.2024.3468992}
}

@article{moreno2017,
	title        = {{Planning Low-Carbon Electricity Systems Under Uncertainty Considering Operational Flexibility and Smart Grid Technologies}},
	author       = {R. Moreno and A. Street and J. M. Arroyo and others},
	year         = 2017,
	journal      = {Philosophical Transactions of the Royal Society A: Mathematical, Physical and Engineering Sciences},
	volume       = 375,
	number       = 2100,
	doi          = {10.1098/rsta.2016.0305}
}

@article{apablaza2024,
	title        = {{Assessing the Impact of DER on the Expansion of Low-Carbon Power Systems Under Deep Uncertainty}},
	author       = {Apablaza, P. and Puschel-Lovengreen, S. and Moreno, R. and others},
	year         = 2024,
	journal      = {Electric Power Systems Research},
	volume       = 235,
	doi          = {10.1016/j.epsr.2024.110824}
}

@article{flores2016,
	title        = {{A Column Generation Approach for Solving Generation Expansion Planning Problems With High Renewable Energy Penetration}},
	author       = {Angela Flores-Quiroz and Rodrigo Palma-Behnke and Golbon Zakeri and others},
	year         = 2016,
	journal      = {Electric Power Systems Research},
	volume       = 136,
	pages        = {232--241},
	doi          = {10.1016/j.epsr.2016.02.011}
}

@article{flores2021,
	title        = {{A Distributed Computing Framework for Multi-Stage Stochastic Planning of Renewable Power Systems With Energy Storage as Flexibility Option}},
	author       = {Angela Flores-Quiroz and Kai Strunz},
	year         = 2021,
	journal      = {Applied Energy},
	volume       = 291,
	pages        = 116736,
	doi          = {10.1016/j.apenergy.2021.116736}
}

@article{jia2019,
	title        = {{Benders Cut Classification via Support Vector Machines for Solving Two-Stage Stochastic Programs}},
	author       = {Jia, Huiwen and Shen, Siqian},
	year         = 2021,
	journal      = {INFORMS Journal on Optimization},
	volume       = 3,
	number       = 3,
	pages        = {278--297},
	doi          = {10.1287/ijoo.2019.0050}
}

@article{borozan2024,
	title        = {{Machine Learning-Enhanced Benders Decomposition Approach for the Multi-Stage Stochastic Transmission Expansion Planning Problem}},
	author       = {S. Borozan and others},
	year         = 2024,
	journal      = {Electric Power Systems Research},
	volume       = 237,
	doi          = {10.1016/j.epsr.2024.110985}
}

@inproceedings{zamzam2020,
	title        = {{Learning Optimal Solutions for Extremely Fast AC Optimal Power Flow}},
	author       = {Zamzam, Ahmed S. and Baker, Kyri},
	year         = 2020,
	booktitle    = {2020 IEEE International Conference on Communications, Control, and Computing Technologies for Smart Grids (SmartGridComm)},
	pages        = {1--6},
	doi          = {10.1109/SmartGridComm47815.2020.9303008}
}

@book{ordoudis2016updated,
	title        = {{An Updated Version of the IEEE RTS 24-Bus System for Electricity Market and Power System Operation Studies}},
	author       = {Ordoudis, Christos and Pinson, Pierre and Morales, Juan M and others},
	year         = 2016,
	publisher    = {Technical University of Denmark},
	volume       = 13
}

@misc{2024_ISP,
	title        = {{2024 Integrated System Plan ISP}},
	author       = {AEMO},
	year         = 2024,
}

@misc{NER_3.9.3C,
	title        = {{National Electricity Rules, Clause 3.9.3C: Reliability Standard and Interim Reliability Measure}},
	author       = {Australian Energy Market Commission},
	year         = 2020,
}

@article{fast_zhan_2017,
	title        = {{A Fast Solution Method for Stochastic Transmission Expansion Planning}},
	author       = {Zhan, Junpeng and Chung, C.Y. and Zare, Alireza},
	year         = 2017,
	journal      = {IEEE Transactions on Power Systems},
	doi          = {10.1109/tpwrs.2017.2665695}
}

@article{Fischetti2018,
	title        = {{Deep Neural Networks and Mixed Integer Linear Optimization}},
	author       = {Matteo Fischetti and Jason Jo},
	year         = 2018,
	journal      = {Constraints},
	volume       = 23,
	number       = 3,
	pages        = {296--309},
	doi          = {10.1007/s10601-018-9285-6}
}

@misc{dumouchelle2022neur2spneuraltwostagestochastic,
	title        = {{Neur2SP: Neural Two-Stage Stochastic Programming}},
	author       = {Justin Dumouchelle and Rahul Patel and Elias B. Khalil and others},
	year         = 2022,
    howpublished = {arXiv preprint arXiv:2205.12006}
}

@article{IEEE_33_bus,
	title        = {{An Enhanced IEEE 33 Bus Benchmark Test System for Distribution System Studies}},
	author       = {Dolatabadi, Sarineh Hacopian and Ghorbanian, Maedeh and Siano, Pierluigi and others},
	year         = 2021,
	journal      = {IEEE Transactions on Power Systems},
	volume       = 36,
	number       = 3,
	pages        = {2565--2572},
	doi          = {10.1109/tpwrs.2020.3038030}
}

@book{hastie2009elements,
	title        = {{The Elements of Statistical Learning: Data Mining, Inference, and Prediction}},
	author       = {Hastie, Trevor and Tibshirani, Robert and Friedman, Jerome},
	year         = 2009,
	publisher    = {Springer},
	note         = {See Section 10.6 on loss functions and differentiability},
	edition      = {2nd}
}

@article{Grimstad_2019,
   title={ReLU networks as surrogate models in mixed-integer linear programs},
   volume={131},
   journal={Computers \& Chemical Engineering},
   publisher={Elsevier BV},
   author={Grimstad, Bjarne and Andersson, Henrik},
   year={2019}
   }

@INPROCEEDINGS{10267731,
  author={Rossini, M. and Rossi, M. and Siface, D.},
  booktitle={27th International Conference on Electricity Distribution (CIRED 2023)}, 
  title={A surrogate model of distribution networks to support transmission network planning}, 
  year={2023},
}

@article{LEDEE2025125812,
    title = {Improved surrogate modeling for multi-energy system design: Model architecture, sampling and scaling choices},
    journal = {Applied Energy},
    volume = {390},
    pages = {125812},
    year = {2025},
    author = {François Lédée and Curran Crawford and Ralph Evins}
}

@article{spyrou2024,
    title = {{How to Assess Uncertainty-Aware Frameworks for Power System Planning?}},
    year = {2024},
    journal = {IEEE Transactions on Energy Markets, Policy and Regulation},
    author = {Spyrou, Elina and Hobbs, Ben and Chattopadhyay, Deb and Mukhi, Neha},
    month = {2},
    pages = {1--13},
    publisher = {Institute of Electrical and Electronics Engineers (IEEE)},
    doi = {10.1109/tempr.2024.3365977}
}

@incollection{cg_book,
    title = {{Implementing Mixed Integer Column Generation}},
    year = {2005},
    booktitle = {Desaulniers, G., Desrosiers, J., Solomon, M.M. (eds) Column Generation.},
    author = {Vanderbeck, F},
    pages = {331--358},
    publisher = {Springer},
    address = {Boston, MA},
    doi = {http://dx.doi.org/10. 1007/0-387-25486-2{\_}12}
}

@article{philpott2009dantzig,
    title = {{Dantzig-Wolfe Decomposition for Solving Multistage Stochastic Capacity-Planning Problems}},
    year = {2009},
    journal = {Operations Research},
    author = {Singh, Kavinesh J and Philpott, Andy B and Wood, R Kevin},
    number = {5},
    pages = {1271--1286},
    volume = {57},
    doi = {10.1287/opre.1080.0678}
}

@inproceedings{KeimMachineLC,
  title={Machine Learning–Enhanced Column Generation for Large-Scale Capacity Planning Problems},
  author={Felipe Keim and V´ıctor Bucarey and Qi Zhang and {\'A}ngela Flores-Quiroz}
}

@inproceedings{NIPS2017_8a20a862,
 author = {Lundberg, Scott M and Lee, Su-In},
 booktitle = {Advances in Neural Information Processing Systems},
 pages = {},
 title = {A Unified Approach to Interpreting Model Predictions},
 volume = {30},
 year = {2017}
}

\end{document}